\documentclass[12pt]{article}
\usepackage{epsf}

\begin{document}

\title{
{\bf Nonlinearity of Regge trajectories in the scattering region}}
\author{{\it A.A. Godizov\thanks{E-mail: godizov@sirius.ihep.su}, 
V.A. Petrov}\thanks{E-mail: Vladimir.Petrov@ihep.ru}\\
{\small {\it Institute for High Energy Physics, Protvino, Russia}}}
\date{}
\maketitle

\vskip-1.0cm

\begin{abstract}
The nonlinearity of Regge trajectories at real negative values of 
the argument is discussed as their general QCD-inspired property. 
The processes of elastic diffractive scattering $p+p\to p+p$ and 
$\bar p+p\to\bar p+p$ at collision energies $\sqrt{s}>23\,GeV$ and 
transferred momenta squared $0.005\,GeV^2<-t<3\,GeV^2$ are considered 
in the framework of the Regge-eikonal model \cite{arnold}. 
By comparison of phenomenological estimates with available 
experimental data on angular distributions it is demonstrated 
that in this kinematical range the data can be satisfactorily 
described as if taking into account only 
three nonlinear Regge trajectories with vacuum quantum numbers 
(``soft'' pomeron, $C$-even $f_2/a_2$-reggeon 
and $C$-odd $\omega/\rho$-reggeon). It is also 
shown that their nonlinearity is essential and not to be ignored. The 
correspondence of 
the Kwiecinski $q\bar q$-pole \cite{kwiecinski} to the 
secondary reggeons and the relevance of the 
Kirschner-Lipatov ``hard'' 
pomeron pole \cite{kirschner} to elastic diffraction are discussed.
\end{abstract}

\vspace*{1cm}

\section*{Introduction}

Our goal is to substantiate that nonlinearity of leading Regge trajectories 
at negative values of the argument (nonlinearity of trajectories in the 
resonance energy region is discussed in \cite{tang}, \cite{inopin}) 
following from some general requirements is essential and not 
to be neglected under considering strong interaction phenomena in the 
framework of Regge-based models. We will demonstrate this on the example of 
high energy elastic (anti)pp-scattering and will exploit for this purpose 
the Regge-eikonal model \cite{arnold}. 
We do not intend to compete with \cite{landshoff}, \cite{shelkovenko}, 
\cite{martynov}, \cite{povh}, \cite{bourrely}, 
\cite{islam}, \cite{avila}, \cite{martynov2} and other authors 
in getting the lowest value of $\chi^2$/d.o.f. over all available 
data on angular distributions at high energies. 

The nonlinearity of the leading trajectory was demonstrated in the 
experiment on measurement of single diffraction cross-sections \cite{UA8}. 
Also, the use of the effective nonlinear dipole pomeron trajectory 
provided the successful description of existing data on high-energy 
elastic (anti)pp-scattering \cite{shelkovenko}, on photoproduction 
of vector mesons \cite{predazzi}, \cite{fiore} and 
deeply virtual Compton scattering \cite{capua}. 

Note also that authors \cite{landshoff}, \cite{kaidalov} 
as well as many others insist on the linearity of some leading Regge 
trajectories in the Euclidean domain at least at $-2\,GeV^2<t<0$. 
The first phenomenological argument is a natural desire to continue 
Chew-Frautschi plots to the region of negative values of the argument. But 
all poles corresponding to resonances are situated on different unphysical 
sheets since they possess not only the mass but also the nonzero width. 
Hence, appropriate points on any Chew-Frautschi plot pertain to different 
branches of the corresponding analytic function and straight continuation 
to the region below the lowest threshold is not correct from the analytical 
point of view (although it can be used, for example, for rough estimation of 
the intercept value). We have to conclude that hadron spectroscopy does not 
provide absolutely reliable grounds for determining the behavior of 
Regge trajectories in the Euclidean domain. The second ``evidence'' of the 
linearity originates from the data on exchange processes 
$\pi^-+p\to \pi^0+n$ and $\pi^-+p\to \eta+n$ in the Born approximation 
\cite{irving}. Since within this approach only two of the leading reggeons 
($\rho$ and $a_2$) give contribution to the amplitude and we can consider 
these trajectories 
approximately equal (as a consequence of the weak degeneracy following from 
the hadron spectroscopy data) one could try to extract them directly 
from the corresponding high-energy angular distributions 
\cite{barnes}. The appropriate formula is 
\begin{equation}
\label{born}
\alpha(t)=1+\frac{1}{2}\left(\ln\frac{s_1}{s_2}\right)^{-1}
\left(\left.\ln\frac{d\sigma}{dt}\right|_{s=s_1}-
\left.\ln\frac{d\sigma}{dt}\right|_{s=s_2}\right)\,.
\end{equation}
For any reaction and any different collision energies 
$\sqrt{s_1}$ and $\sqrt{s_2}$ the function 
$\alpha(t)$ must be the same (if 
the Born approximation works) and, hence, the 
function $r(t)\equiv(1-\alpha^{\pi^0+n}(t))/(1-\alpha^{\eta+n}(t))$ 
(where $\alpha^{\pi^0+n}(t)$ and $\alpha^{\eta+n}(t)$ are 
the trajectories from 
angular distributions for prosesses $\pi^-+p\to \pi^0+n$ and 
$\pi^-+p\to \eta+n$ correspondingly at collision energies 
$\sqrt{s_1}=11.0\,GeV$ and $\sqrt{s_2}=19.4\,GeV$) must be 
strictly equal to unity at any 
argument value. 
However, this is not evident after extraction this function from the 
data on exchange reactions (fig. below). 
\begin{figure}[h]
\hskip 5cm
\epsfxsize=8cm\epsfysize=5cm\epsffile{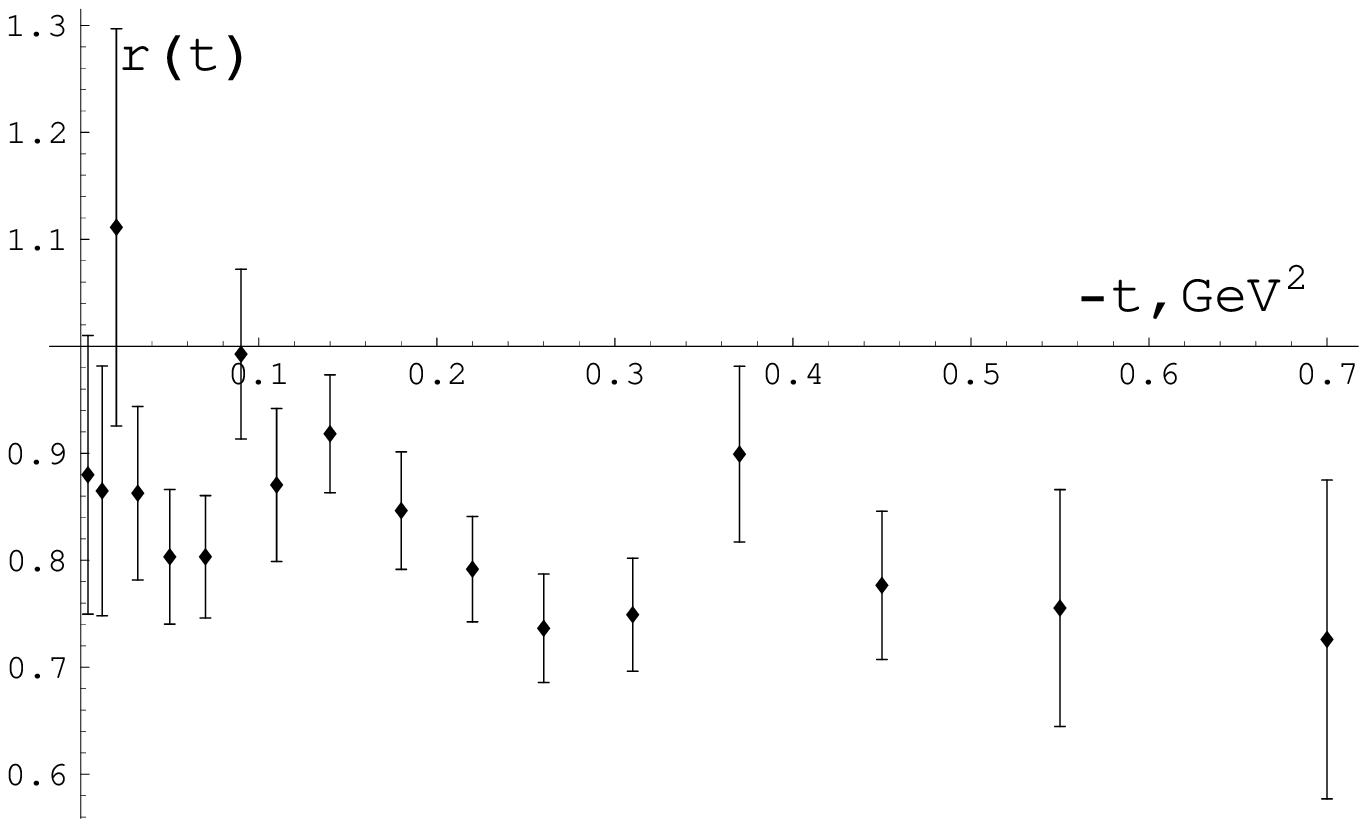}
\end{figure}
The only conclusion which an unprejudiced person can draw looking at this 
picture is that the Born approximation seems invalid at transferred 
momenta $-t>0.2\,GeV^2$ and, consequently, the linearity of 
the $\rho/a_2$-trajectory at higher scattering angles 
is not guaranteed. 

So, linear parametrizations for Regge trajectories have no 
phenomenological advantages over nonlinear ones. 
Also note that the best fit 
to the trajectory extracted from the earlier data on $\pi^-+p\to \eta+n$ 
\cite{guisan} under exploiting the Born approximation was obtained for 
nonlinear parametrization \cite{squires}, \cite{philips}
\begin{equation}
\label{pignotti}
\alpha(t)=\alpha(\infty)+\frac{[\alpha(0)-\alpha(\infty)]^2}
{\alpha(0)-\alpha(\infty)-t\alpha'(0)}
\end{equation}
with $\alpha(\infty)=0$. 

Our viewpoint is based on the conviction that 
QCD is the fundamental theory of strong interaction. This rather 
general requirement imposes restrictions on the behavior of Regge 
trajectories in the range of the perturbative QCD validity since 
at large transferred momenta exchanges by single 
reggeons must turn into exchanges by colour-singlet parton combinations. 
If we assume that the reggeon exchange giving the leading contribution 
to the eikonal at high energies (pomeron) turns into some multi-gluon 
exchange in the perturbative range we will come to \cite{wu}, \cite{low}, 
\cite{kearney}, \cite{kirschner} 
\begin{equation}
\label{gluon}
\lim_{t\to -\infty}\alpha_{gg...g}(t) = 1\,.
\end{equation}
In the case of the quark-antiquark pair ($f_2$-reggeon, $\omega$-reggeon 
etc.) one obtains \cite{kwiecinski} 
\begin{equation}
\label{meson}
\alpha_{\bar qq}(t) = 
\sqrt{\frac{8}{3\pi}\alpha_s(\sqrt{-t})}+o(\alpha_s^{1/2}(\sqrt{-t}))
\end{equation}
where $\alpha_s(\mu)\equiv g_s^2(\mu)/4\pi$ is the running 
coupling.\footnote{The universality of the 
asymptotic vanishing at $t=-\infty$ of meson trajectories seems 
to contradict the existence of (pseudo)scalars. In fact, if $\alpha_\pi(t)$ 
behaved like (\ref{meson}) at $t\to -\infty$ it could not be monotonic 
function of $t$. Possible way to preserve the monotony is to suppose that 
at $t\to -\infty$ the trajectory tends to some negative value. Similar 
behavior takes place, e.g., in the phenomenological studies of the trajectories 
containing heavy quarkonia \cite{likhoded}. To what extent is it possible 
in QCD remains unclear.} 

This property of Regge poles (tending to constant at $t\to -\infty$) is 
quite general and follows 
from the fact of their invariance relative to the renormalization 
group transformations (the requirement of renorm-invariance is well-grounded 
by observability of bound states and resonances). 
The general solution of the renorm-group differential 
equation for the renorm-invariant quantity $f(\frac{t}{\mu^2}\,,\;\alpha_s(\mu))$ 
is (in the case of massless fields) of the form \cite{shirkov} 
\begin{equation}
\label{renormdec}
f\left(\frac{t}{\mu^2}\,,\;\alpha_s(\mu)\right) = 
\Phi\left(\frac{t}{\mu^2}e^{K(\alpha_s(\mu))}\right)
\end{equation}
where $\Phi(x)$ is an arbitrary function analytic in the region defined by the 
analyticity in $t$, $\mu$ is the renormalization scale, 
$K'(\alpha_s) = 1/\beta(\alpha_s)
\equiv (\mu^2\frac{\partial \alpha_s(\mu)}{\partial\mu^2})^{-1}$. 
For any quantum field model with asymptotic freedom 
(\ref{renormdec}) leads in the perturbative sector 
($\mu = \sqrt{-t}$, $t\to -\infty$) to 
\begin{equation}
\label{renormasy}
\lim_{t\to -\infty}f(t) = const\,.
\end{equation}

Actually, this corresponds to the existence of the free field limit 
as $\alpha_s(\sqrt{-t})\to 0$. In the theory of potential scattering the 
squared effective radius of interaction corresponding to the reggeon exchange 
\cite{squires}, \cite{regge} $R^2\sim \alpha'(t)(2\alpha(t)+1)\to 0$ 
at $t\to -\infty$, $\alpha(t)\to const$. This purely quantum mechanical 
result is in agreement with the property of asymptotic freedom at short 
distances (in the case 
of linear trajectory $\alpha(t)$ ($\alpha'(t)>0$) $R^2\to -\infty$ at 
$t\to -\infty$). 

As a consequence of (\ref{renormasy}) the essential 
nonlinearity of the Regge trajectories takes place in the range $-\infty<t<0$. 
This nonlinearity is their fundamental property. 

Besides, we assume that $Im\,\alpha(t+i0)\ge 0$ 
increases slowly enough at 
$t\to +\infty$ (for example, not faster than $Ct\ln^{-1-\epsilon}t$, 
$\epsilon>0$) so that the dispersion relations with 
not more than one subtraction take place, i.e. 
\begin{equation}
\label{disper}
\alpha(t)=\alpha_0+\frac{t}{\pi}\int_{t_T}^{+\infty}
\frac{Im\,\alpha(t'+i0)}{t'(t'-t)}dt'\,,
\end{equation}
and $Im\,\alpha(t+i0)\ge 0$ at $t\ge t_T>0$ 
(we would like to point out that these assumptions are strictly 
fulfilled in the theory of perturbations and the theory of 
potential scattering \cite{collins}). In this 
case we obtain \cite{squires}, \cite{collins}
\begin{equation}
\label{gerg}
\frac{d^n\alpha(t)}{dt^n}>0\;\;(t<t_T,\;n=1,2,3,...).
\end{equation}

The $t$-channel unitarity implies some special $t$-dependence of the 
trajectories near the threshold \cite{collins} 
$$
Im\,\alpha(t)\sim(t-t_T)^{\alpha(t_T)+1/2}\;\;\;(t\ge t_T)\,.
$$
As was argued in \cite{cohen} this must have observable effects. 
However, in this paper we do not take this requirement into account, 
considering such effects as a kind of ``fine structure'' which is beyond 
the accuracy level we adopted.

It is shown in \cite{prokudin} that the assumption of the linearity of 
Regge trajectories 
leads to the situation when the diffractive pattern at high energy 
(anti)proton-proton elastic scattering can be described in the framework of 
the Regge-eikonal approach only after introduction of several vacuum 
reggeons giving contributions to the eikonal (one needs not less than three 
$C$-even poles with intercepts higher than unity). In other Regge-based 
models one also need more than one reggeon with intercept 
higher than unity (see, for example, \cite{shelkovenko}, \cite{martynov}, 
\cite{martynov2}).
The use of nonlinear trajectories allows us to curtail the number of 
reggeons essential for an 
acceptable description of the data on high energy elastic nucleon-nucleon 
scattering.
Namely, using only one trajectory with vacuum quantum numbers and 
intercept higher than unity (``soft'' pomeron) and two secondary reggeons 
with intercepts lower than unity ($C$-even $f_2/a_2$-reggeon 
and $C$-odd $\omega/\rho$-reggeon) 
we will be able to describe available experimental data 
in a wide kinematical range 
$\sqrt{s} > 23\,GeV$, $0.005\,GeV^2 < -t < 3\,GeV^2$ 
(at $-t>3\,GeV^2$ we can not ignore the contributions from 
other leading vacuum reggeons -- $C$-even ``hard'' pomeron(s) and 
$C$-odd odderon(s)). 
The very fact that we have managed to satisfactorily 
reproduce the diffractive pattern within this rather wide kinematical 
range in the framework of such a simple phenomenological scheme is 
quite encouraging and points to the agreement between general 
QCD-related theoretical conclusions and experiment.

\section*{The QCD pomeron}

Regge trajectories which are tightly related to the hadron spectroscopy 
deal, generally, with confinement of quarks and gluons. Thus, without a 
great progress in the 
solution of this outstanding problem the QCD theory of Regge trajectories 
is still in its primordial stage. 

It is natural that in the absence of a regular non-perturbative technique 
some progress is limited by perturbative calculations. In the literature 
the wide attention is being paid to so-called ``BFKL pomeron'' (or other 
``BFKL reggeons'') \cite{kuraev}. In this approach one strives to 
formulate some kind of Bethe-Salpeter equation for the gluon-gluon 
scattering amplitude (Green function) in which ``$t$-channel'' gluons are 
actually gluon Regge trajectories (``reggeized gluons'') that have to be 
preliminary calculated from another Bethe-Salpeter equation for the 
colour-octet $t$-channel. This trick can make an impression that, say, the 
pomeron trajectory is not just a 2-gluon exchange (with account of interaction 
between these two gluons via ``ladder rung exchanges'') but contains 
many-gluon configurations in the $t$-channel as well. 
However, the very ``reggeized gluons'' are related to no more than two 
gluons each (since they result from the the BS-type equation with
2-gluon irreducible kernels). That is one deals with no more than 4 
usual gluons in the $t$-channel. 

The use of fixed number of gluon exchanges seems to be justified at $|t|$ 
large enough, where the elementary short-distance structure of reggeons 
has to show up. As to small values of $t$ it is clear that due to genuinely 
strong interaction one cannot limit the problem by any fixed number 
of exchanged partons. Certainly, one can argue about ``valence'' gluons but 
these have to be essentially non-perturbative and different from the ``slightly 
reggeized'' gluons mentioned above. It is even quite probable that the 
corpuscular language in this quasi-classical region seases to be adequate and 
undulatory gluonic fields are more relevant. 

In spite of such a little bit gloomy landscape one still can try to account 
(when describing the scattering data in terms of Regge exchanges) for such a 
distinctive QCD prediction as asymptotic constancy of Regge trajectories 
in deeply Euclidean region, $-t\to\infty$ (in other words, the nonlinearity 
of the pomeron and secondary reggeon trajectories which was already mentioned 
in the previous section). Via solving the Bethe-Salpeter-like equation 
for the gluon-gluon scattering amplitude 
it was obtained in \cite{kirschner} that 
\begin{equation}
\label{hard}
\alpha^{hard}_P(t) = 1+\frac{12\,\ln 2}{\pi}\alpha_s(\sqrt{-t})
\left[1-\alpha_s^{2/3}(\sqrt{-t})\left(\frac{7\zeta(3)}{2\,\ln 2}\right)^{1/3}
\left(\frac{3/4}{11-2/3\,n_f}\right)^{2/3}\right]\,,\;\;\;t\to -\infty\,.
\end{equation}

The second term in brackets is approximately equal to $0.09$ at $t=-M_Z^2$ 
when $\alpha_s(M_Z)\approx 0.118$ 
(here $M_Z\approx 91.2\,GeV$ is the Z-boson mass and $n_f=5$) 
and such an expansion is justified 
at this scale (at $t=-(2.5\,GeV)^2$, $\alpha_s(2.5\,GeV)\approx 0.3$ this term is 
about 0.17). Even at $t=-M_Z^2$ 
the value of $\alpha^{hard}_P(-M_Z^2)\approx 1.28$ is quite high 
($\alpha^{hard}_P(-6\,GeV^2)\approx 1.66$). If one assumes the monotony 
of $\alpha^{hard}_P(t)$ the intercept, $\alpha^{hard}_P(0)$, 
has to lie even higher. The last statement agrees with a rough estimate of the 
lower bound for the BFKL pomeron intercept value 
$\alpha^{hard}_P(0)-1\ge 0.3$ obtained in \cite{lipatov}. 
Such a value can be hardly relevant for the data on elastic diffraction. 

Some time ago a NLO result was obtained for the pomeron intercept \cite{camici} 
\begin{equation}
\label{hardinter}
\alpha_P(0) = 1+\frac{12\,\ln 2}{\pi}\alpha_s(\mu)
\left(1-\frac{20}{\pi}\alpha_s(\mu)\right)\,.
\end{equation}
This expression cannot be accepted as a true value of the pomeron intercept 
as it depends on the renormalization scheme and arbitrary renormalizaton scale 
$\mu$ via $\alpha_s$. As was argued recently \cite{petrov} within the limit 
of massless quark fields the true intercept of 
any Regge trajectory has to be strictly independent of the QCD coupling 
constant. 

These curcumstances enforce us (while waiting for further theoretical 
progress) to try some purely phenomenological 
ansatz for the pomeron trajectory which retains but one feature of 
(\ref{hard}): it tends to 1 at $|t|$ high enough (see relation (\ref{gluon})). 

\section*{The model}

Since our primary goal is not to get the best phenomenological 
description of all available data on elastic diffraction 
at high energies but to find a phenomenological confirmation of 
such a macroscopic phenomenon 
as following from QCD nonlinear behavior of Regge trajectories 
in the scattering region we choose the 
eikonal to have the simplest form (ignoring the contribution of 
``hard'' pomeron(s), odderon(s) etc.)
\begin{equation}
\label{eikphen}
\delta(s,t) = \delta_P(s,t)+\delta_+(s,t)\mp\delta_-(s,t)=
\left(i+tg\frac{\pi(\alpha_P(t)-1)}{2}\right)
\beta_P(t)\left(\frac{s}{s_0}\right)^{\alpha_P(t)}+
$$
$$
+\left(i+tg\frac{\pi(\alpha_+(t)-1)}{2}\right)
\beta_+(t)\left(\frac{s}{s_0}\right)^{\alpha_+(t)}\mp
\left(i-ctg\frac{\pi(\alpha_-(t)-1)}{2}\right)
\beta_-(t)\left(\frac{s}{s_0}\right)^{\alpha_-(t)}
\end{equation}
where $s_0\equiv 1\,GeV^2$, $\alpha_P(t)$ and $\beta_P(t)$ are the 
trajectory and the residue of the ``soft'' pomeron, 
$\alpha_+(t)$ and $\alpha_-(t)$ are the trajectories of secondary 
reggeons (i.e. we assume that $\alpha_+(t)\approx \alpha_{f_2}(t)
\approx \alpha_{a_2}(t)$ and $\alpha_-(t)\approx \alpha_\omega(t)
\approx \alpha_\rho(t)$ due to the isospin symmetry of the quark flavors), 
$\beta_+(t)\equiv\beta_{f_2}(t)+\beta_{a_2}(t)$ and 
$\beta_-(t)\equiv\beta_\omega(t)+\beta_\rho(t)$, the sign ``--'' (``+'')
before $\delta_-(t)$ 
corresponds to the particle-on-particle (particle-on-antiparticle) 
scattering.

Our phenomenological ansatz for the ``soft'' 
pomeron pole function is (see fig. below)
\begin{equation}
\label{pomeron}
\alpha_P(t) = 1+p_1\left[1-p_2\,t\left(arctg(p_3-p_2\,t)
                             -\frac{\pi}{2}\right)\right]\,.
\end{equation}
\begin{figure}[h]
\hskip 5cm
\epsfxsize=8cm\epsfysize=4.5cm\epsffile{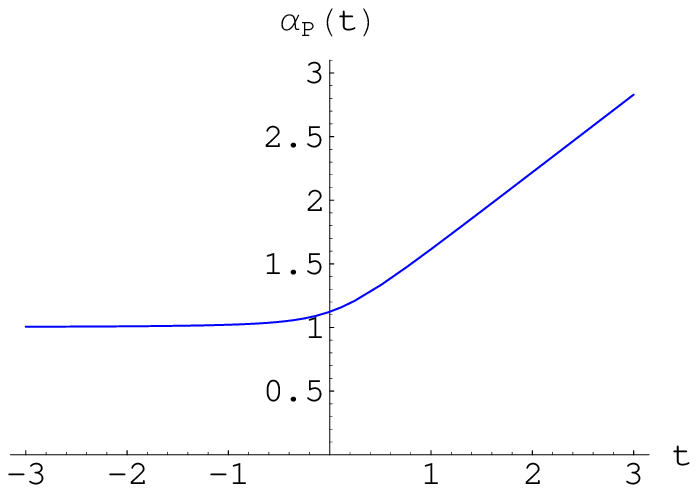}
\end{figure}
It bears the above-mentioned characteristic feature, i.e. tends to 1 at 
$t\to -\infty$. At large and positive $t$ (\ref{pomeron}) mimicrizes the 
``stringy'' behavior, i.e. grows linearly with $t$. We do not give a great 
significance to this, though. Moreover, such a behavior, if taken seriously, 
leads to complex singularities of $\alpha_P(t)$ and one has to provide a 
special care to save the amplitude from causality (analyticity) violations
\cite{thorn}.\footnote{Earlier, 
in \cite{ilyin} there was proposed a similar form for the 
pomeron trajectory $\alpha_P(t)=1+\alpha_P'(0)s_1\,arctg\frac{t}{s_1}$. This 
function also flattens at $t\to -\infty$ but in \cite{ilyin} $s_1$ was 
chosen large enough for the linear approximation was valid in the 
diffraction region. As a consequence, 
$\lim_{t\to -\infty}\alpha_P(t)=1-\alpha_P'(0)s_1\frac{\pi}{2}<0$. So, 
our approach (\ref{pomeron}) is quite different because (as it will 
be shown below) asymptotic relation (\ref{gluon}) together with the 
requirement of the 
monotony of $\alpha_P(t)$ results in a strong nonlinearity 
of the pomeron trajectory.}

Residues are chosen as 
\begin{equation}
\label{resid}
\beta_P(t) = B_Pe^{b_P\,t}(1+d_1\,t+d_2\,t^2+d_3\,t^3+d_4\,t^4)\,,\;\;\;\;
\beta_+(t) = B_+e^{b_+\,t},\;\;\;\;\beta_-(t) = B_-e^{b_-\,t}\,.
\end{equation}
Our parametrization for the secondary trajectories contains the 
QCD-inspired expressions 
\begin{equation}
\label{fmeson}
\alpha_+(t) = \left(\frac{8}{3\pi}
\gamma(\sqrt{-t+c_+})\right)^{1/2},\;\;\;\;
\alpha_-(t) = \left(\frac{8}{3\pi}\gamma(\sqrt{-t+c_-})\right)^{1/2}
\end{equation}
where 
\begin{equation}
\label{analytic}
\gamma(\mu) \equiv \frac{4\pi}{11-\frac{2}{3}n_f}
\left(\frac{1}{\ln\frac{\mu^2}{\Lambda^2}}
+\frac{1}{1-\frac{\mu^2}{\Lambda^2}}\right)
\end{equation}
is the so-called one-loop analytical 
QCD effective coupling constant \cite{solovtsov}, 
$n_f = 3$ is the number of quark flavors taken into account, 
$\Lambda = \Lambda^{(3)} = 0.346\,GeV$ is the QCD dimensional parameter 
(the value was taken from \cite{bethke}) 
and $c_+,c_->0$ are free phenomenological parameters.

This analytic approximation for the secondary reggeons is obtained in the 
following way. We take an expression for the $q\bar q$ Regge pole 
in the perturbative sector \cite{kwiecinski} derived via solving the 
Bethe-Salpeter-like equation in the range of the perturbative 
QCD validity and then replace the perturbative effective coupling 
constant to expression (\ref{analytic}) obtained in 
the framework of the dispersive approach \cite{solovtsov}. Further, 
the introduction of the free parameters $c_+$ and $c_-$ is the simplest 
phenomenological way to take into account the 
disparity between $\alpha_+(t)$ and $\alpha_-(t)$ 
and not to spoil the asymptotic behavior of 
these trajectories in the perturbative sector 
($-t>>1\,GeV^2$).

To obtain angular distributions we substitute (\ref{pomeron}), 
(\ref{resid}), (\ref{fmeson}) into (\ref{eikphen}), proceed via 
Fourier-Bessel transformation 
\begin{equation}
\label{fourier1}
\delta(s,b) = \frac{1}{16\pi s}\int_0^{\infty}d(-t)J_0(b\sqrt{-t})\delta(s,t)
\end{equation}
to the coordinate representation, using the eikonal representation of the 
scattering amplitude
\begin{equation}
\label{eikrepr}
T(s,b) = \frac{e^{2i\delta(s,b)}-1}{2i}
\end{equation}
through inverse Fourier-Bessel transformation 
\begin{equation}
\label{fourier2}
T(s,t) = 4\pi s\int_0^{\infty}db^2J_0(b\sqrt{-t})T(s,b)
\end{equation}
obtain its value 
in the momentum representation (during numerical calculating integrals from 
(\ref{fourier1}), (\ref{fourier2}) we change upper limits of integration 
to $4\,GeV^2$ and 
$400\,GeV^{-2}\approx(4\,Fm)^2$ correspondingly) and substitute it into the 
expression for the differential cross-section
\begin{equation}
\label{diffsech}
\frac{d\sigma}{dt} = \frac{|T(s,t)|^2}{16\pi s^2}\,.
\end{equation}

In the conclusion of this section we must note that our consideration 
of the high-$|t|$ behavior provides the explanation of their nonlinearity 
in the framework of QCD but we do not intend to say that our approach 
is capable to give the description of cross-sections at fixed angles. 
Our concrete parametrizations for the unknown 
functions $\alpha_P(t)$, $\alpha_+(t)$, $\alpha_-(t)$, $\beta_P(t)$, 
$\beta_+(t)$, $\beta_-(t)$ in the eikonal (\ref{eikphen}) are purely 
phenomenological quantitative approximations valid in the 
soft diffraction region (only relations (\ref{gluon}), (\ref{meson}), 
(\ref{gerg}) have physical sense). Nonetheless, the QCD asymptotics of 
the trajectories (tending to constants at $t\to -\infty$) are fairly 
compatible with, e.g. ``quark counting rules'' \cite{kearney}. 
The dual amplitude with Mandelstam analyticity model (DAMA) 
with logarithmic trajectories \cite{fiore2} provides an alternative 
way to reproduce the QCD-type behavior at fixed angles. 

\section*{The fitting results}

\begin{figure}
\epsfxsize=16.7cm\epsfysize=16.7cm\epsffile{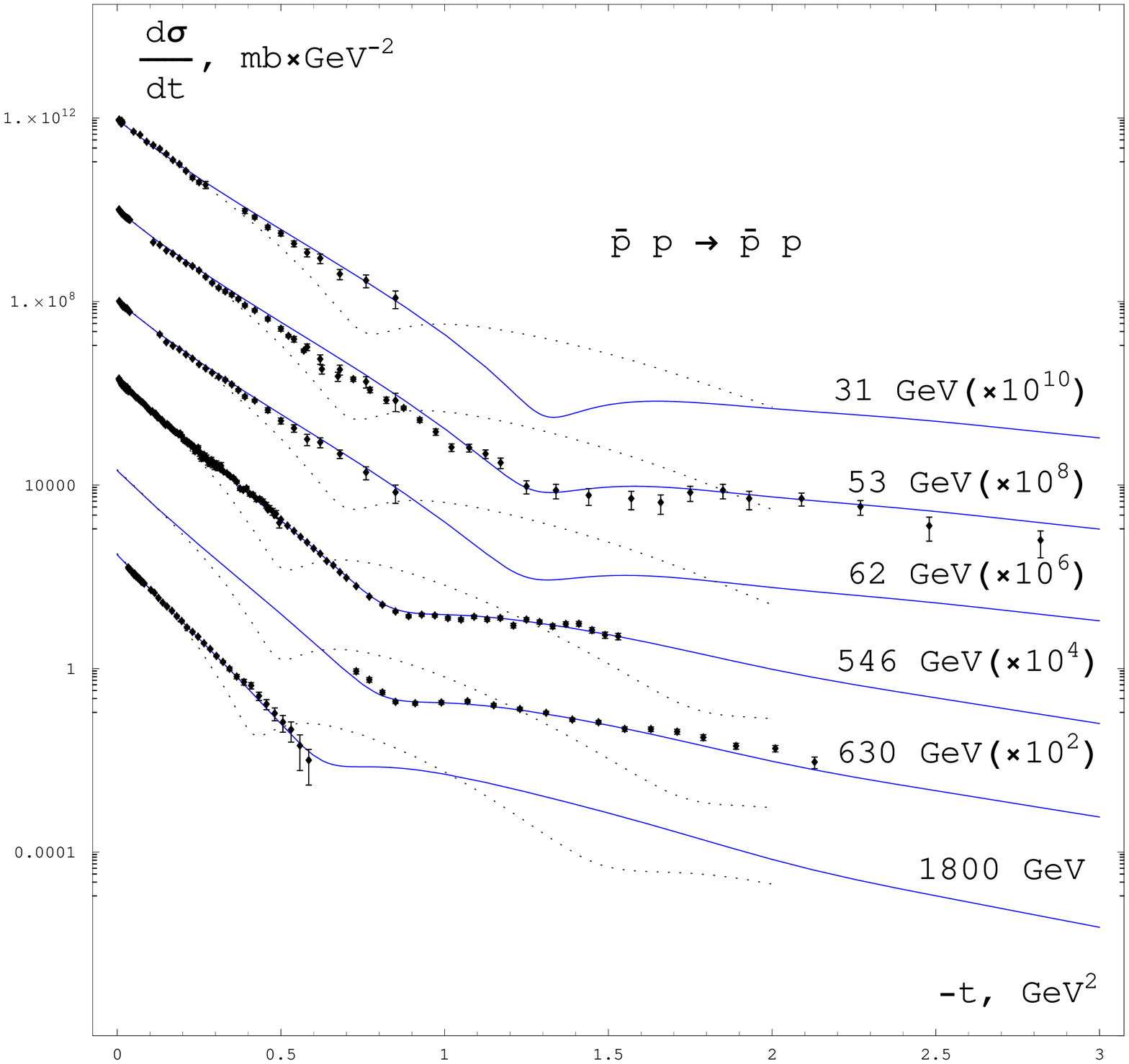}
\caption{Differential $\bar p\;p\to\bar p\;p$ 
cross-sections for the cases of nonlinear (solid lines) and 
linearly approximated (dotted lines) ``soft'' pomeron trajectory.}
\label{app}
\end{figure}

\begin{figure}
\epsfxsize=16.7cm\epsfysize=16.7cm\epsffile{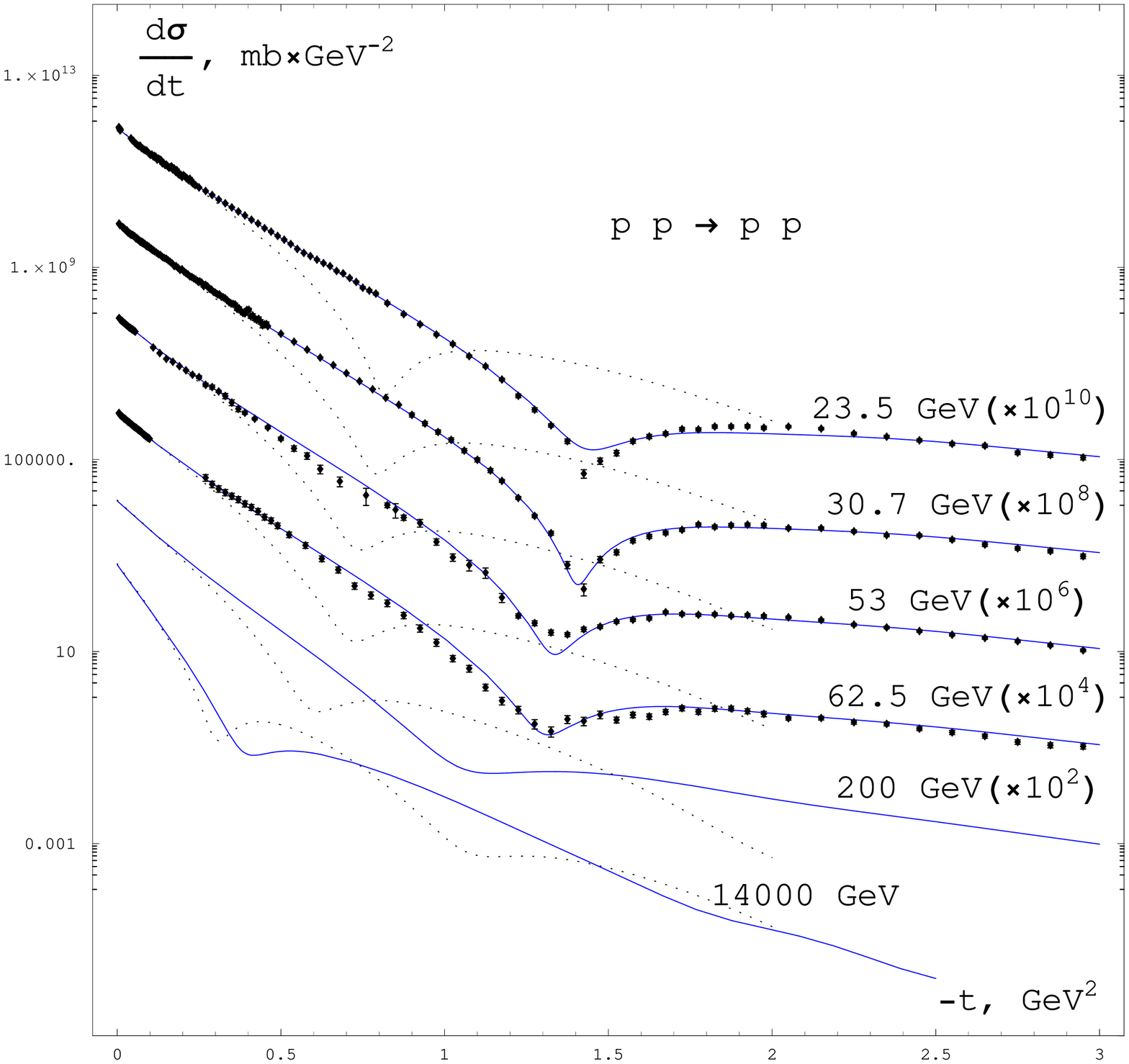}
\caption{Differential $p\;p\to p\;p$ 
cross-sections for the cases of nonlinear (solid lines) and 
linearly approximated (dotted lines) ``soft'' pomeron trajectory.}
\label{pp}
\end{figure}

Turn to the description of experimental data. 
The results of fitting over data on angular distributions 
in the kinematical region $\sqrt{s}>23\,GeV$, 
$0.005\,GeV^2<-t<3\,GeV^2$ 
\cite{diffexp}\footnote{For calculation of 
electromagnetic correstions to the scattering amplitude we used 
the recipe by R. Cahn \cite{cahn} (see also \cite{prokudin2}).} 
are represented in tab. \ref{tab1}, \ref{tab2} 
and fig. \ref{app}, \ref{pp}. 

\begin{table}[h]
\begin{center}
\begin{tabular}{|l|l|l|l|l|l|}
\hline
& {\bf pomeron}  &    & {\bf $f_2/a_2$-reggeon} &  & {\bf $\omega/\rho$-reggeon} \\
\hline
$p_1$ & $0.123$             & $c_+$ & $0.1\,GeV^2$       & $c_-$ & $0.9\,GeV^2$  \\
$p_2$ & $1.58\,GeV^{-2}$    &       &                    &       &               \\
$p_3$ & $0.15$             &       &                    &       &              \\
$B_P$ & $43.5$              & $B_+$  & $153$              & $B_-$ & $46$        \\
$b_P$ & $2.4\,GeV^{-2}$    & $b_+$  & $4.7\,GeV^{-2}$   & $b_-$ & $5.6\,GeV^{-2}$\\
$d_1$ & $0.43\,GeV^{-2}$   &       &                    &       &               \\
$d_2$ & $0.39\,GeV^{-4}$   &       &                    &       &               \\
$d_3$ & $0.051\,GeV^{-6}$   &       &                    &       &              \\
$d_4$ & $0.035\,GeV^{-8}$   &       &                    &       &              \\
\hline
$\alpha_P(0)$ & $1.123$              & $\alpha_+(0)$ & $0.78$             
& $\alpha_-(0)$ & $0.64$ \\
$\alpha'_P(0)$ & $0.28\,GeV^{-2}$   & $\alpha'_+(0)$ & $0.63\,GeV^{-2}$
& $\alpha'_-(0)$ & $0.07\,GeV^{-2}$\\
\hline
\end{tabular}
\end{center}
\caption{Parameters obtained by fitting to the data.}
\label{tab1}
\end{table}

\begin{figure}
\epsfxsize=16.7cm\epsfysize=16.7cm\epsffile{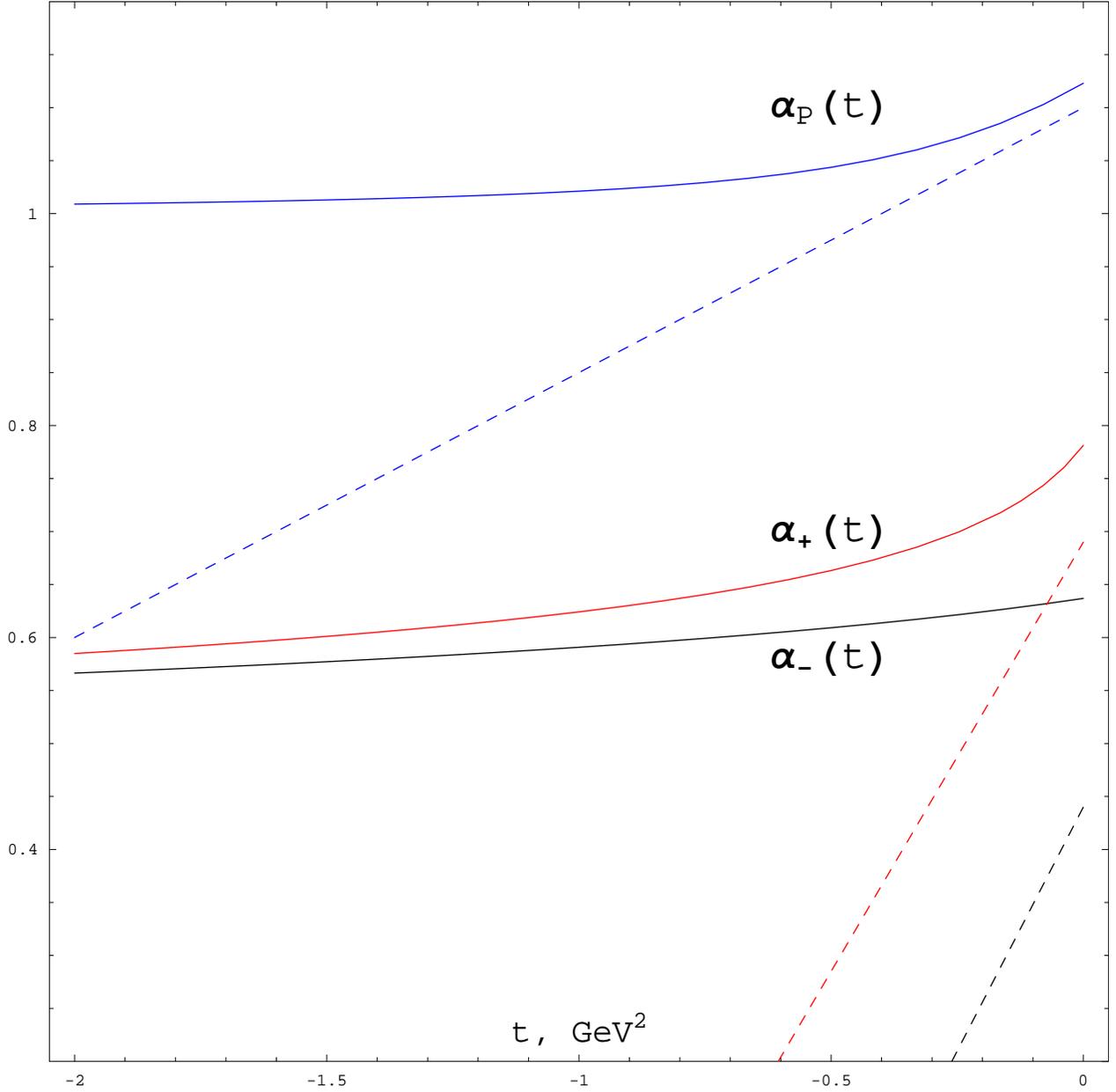}
\caption{Approximate ``soft'' pomeron and secondary trajectories 
obtained in the 
fitting procedure (dashed lines: $\alpha^{lin}_f(t)=0.69+0.81\,t$ 
and $\alpha^{lin}_\omega(t)=0.44+0.92\,t$ are 
the continuations of the Chew-Frautschi plots corresponding to 
$f_2$-reggeon and $\omega$-reggeon and 
$\alpha^{lin}_P(t)=1.1+0.25\,t$ is the linear ``soft'' pomeron 
trajectory usually used in literature).}
\label{tra}
\end{figure}

\begin{figure}
\epsfxsize=16.7cm\epsfysize=16.7cm\epsffile{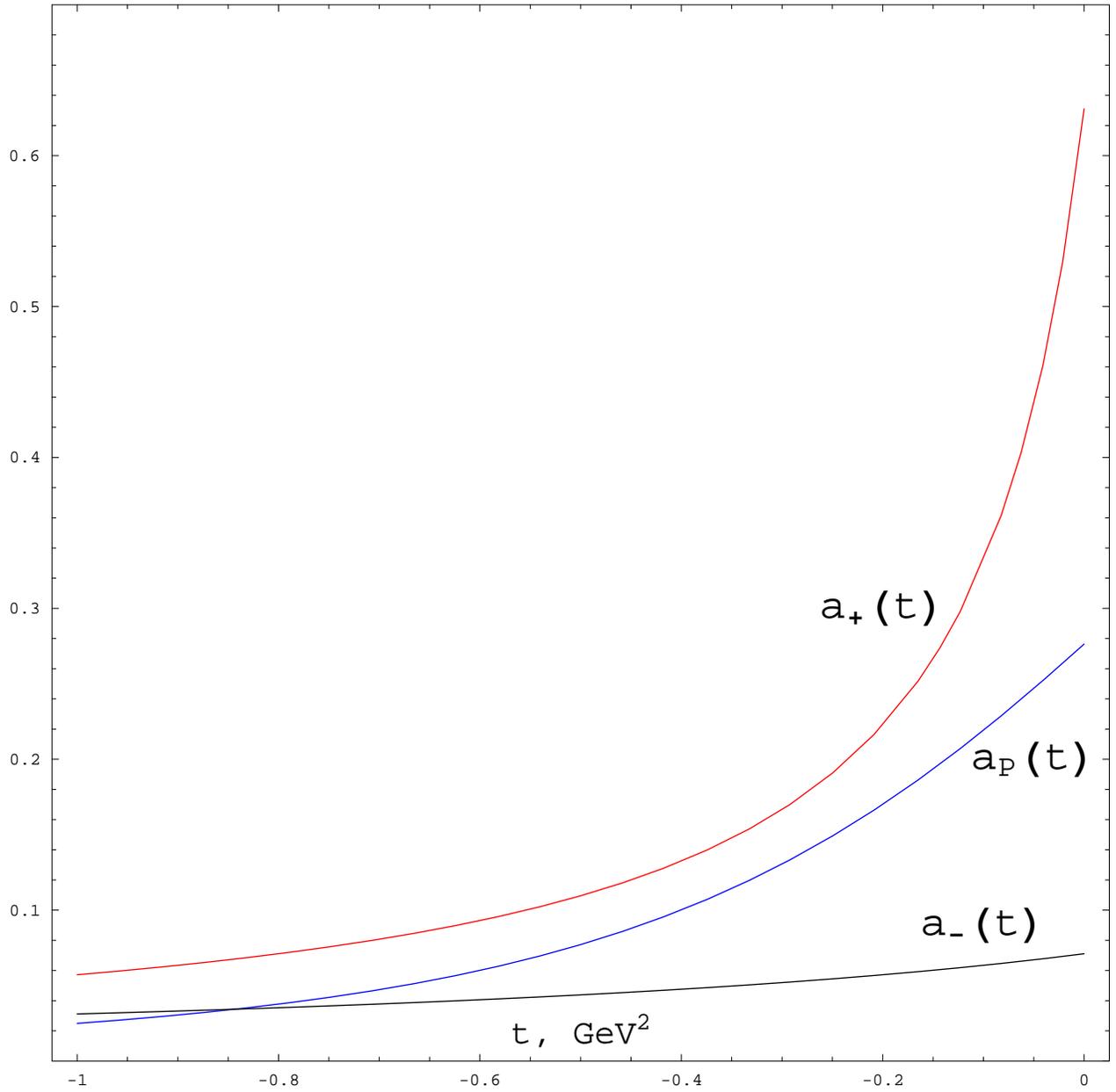}
\caption{The slopes (in $GeV^{-2}$) of the Regge trajectories from 
the previous figure, $a_R(t)\equiv\alpha'_R(t)$.}
\label{slo}
\end{figure}

If we compare angular distributions obtained using parametrization 
(\ref{pomeron}), (\ref{resid}), (\ref{fmeson}) with those obtained using the 
same values of free parameters 
but with replacement of the nonlinear
pomeron trajectory to its linear 
approximation (only two first terms in the Taylor expansion) we will disclose 
a huge difference between the corresponding results 
(see fig. \ref{app}, \ref{pp}) which is a consequence of the fact that the 
simultaneous fulfilment of the conditions 
$\alpha'_P(t)>0$ at $t\le 0$, $\lim_{t\to -\infty}\alpha_P(t) = 1$
together with the phenomenological estimates $\alpha'_P(0)>0.2\,GeV^{-2}$, 
$\alpha_P(0)<0.15$ makes the approximation 
$\alpha_P(t)=\alpha_P(0)+\alpha'_P(0)t$ 
for the pomeron trajectory invalid 
in the region $-t>0.8\,GeV^2$, i.e. the effect of 
the nonlinearity turns out strong even at low $-t$ 
(see fig. \ref{tra}, \ref{slo}). 
Here we must emphasize that the linear approximations to our nonlinear 
trajectories differ from linear trajectories used by other authors.
In the most of papers quoted in tab. \ref{tab3} 
linear Regge trajectories were successfully 
applied to the phenomenological description of the data but QCD 
asymptotic relations (\ref{gluon}), (\ref{meson}) were ignored.

\begin{figure}
\epsfxsize=16.7cm\epsfysize=16.7cm\epsffile{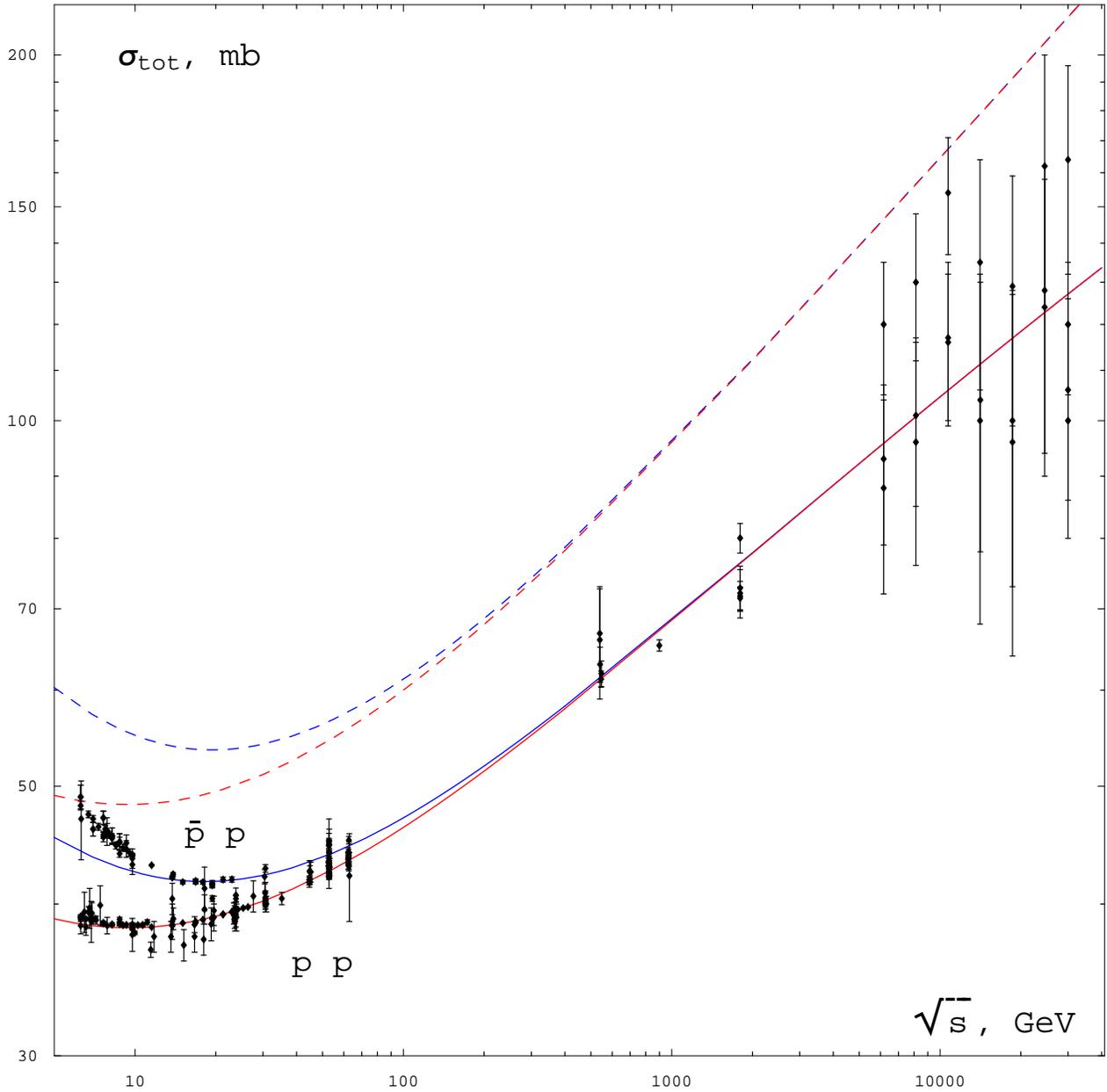}
\caption{Total cross-sections (the eikonalized amplitudes --- solid lines, 
the Born amplitudes --- dashed lines) for high-energy nucleon-nucleon 
scattering as functions of center-of-mass energy (experimental data were 
taken from Particle Physics Data System 
http://wwwppds.ihep.su:8001/ppds.html).}
\label{tot}
\end{figure}

\begin{figure}
\epsfxsize=16.7cm\epsfysize=16.7cm\epsffile{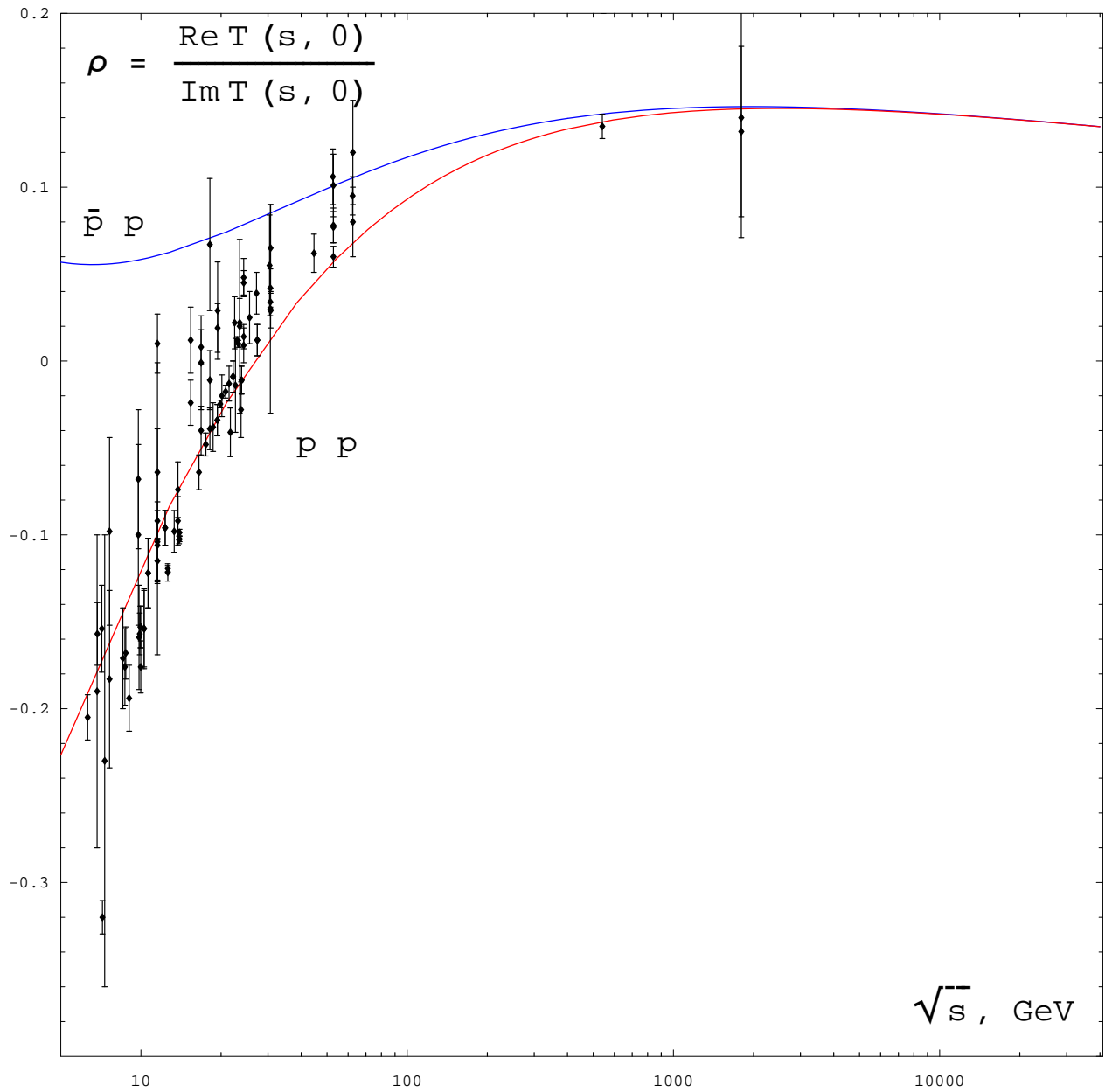}
\caption{The forward amplitude real part to imaginary part ratio 
as a function of center-of-mass energy (experimental data were 
taken from Particle Physics Data System 
http://wwwppds.ihep.su:8001/ppds.html).}
\label{rho}
\end{figure}

In fig. \ref{tot} the predictions for the total cross-section dependence 
on the center-of-mass energy are shown. In particular, 
$\sigma_{tot}(200\,GeV)\approx 52\,mb$, 
$\sigma_{tot}(14\,TeV)\approx 111\,mb$. The noticeable disagreement with 
the experimental $\bar p\,p$ data at $\sqrt{s}<13\,GeV$ 
points to the fact that at such energies the simplest phenomenological 
scheme (\ref{eikphen}) is inapplicable for satisfactory quantitative 
description of the data and the contribution of $f'_2$- and $\phi$-reggeons 
must be taken into account. Since the corresponding terms in the imaginary 
part of the eikonal have equal (opposite) signs for the $\bar p\,p$ 
($p\,p$) scattering the discrepancy between the model curve and the 
$\bar p\,p$ data turns out larger than for the $p\,p$ data. 
Also, in this figure by comparison 
of the imaginary part of the forward amplitude with the same from the 
Born approximation it is demonstrated that we can not neglect absorptive 
corrections which significantly reduce the amplitude value. 
Fig. \ref{rho} represents the dependence of $\rho$-parameter 
on the center-of-mass energy.

\begin{table}[h]
\begin{center}
\begin{tabular}{|l|l|l|}
\hline
Set of data & Number of points &  $\chi^2$               \\
\hline
$\sqrt{s}=23\,GeV$ ($p\,p$)   & 124    & 280\\
$\sqrt{s}=31\,GeV$ ($p\,p$)   & 154    & 467\\
$\sqrt{s}=53\,GeV$  ($p\,p$)    & 85    & 423\\
$\sqrt{s}=62\,GeV$ ($p\,p$)   & 107    & 409\\
$\sqrt{s}=31\,GeV$ ($\bar p\,p$)   & 38    & 108\\
$\sqrt{s}=53\,GeV$ ($\bar p\,p$)   & 60    & 336\\
$\sqrt{s}=62\,GeV$ ($\bar p\,p$)   & 40    & 156\\
$\sqrt{s}=546\,GeV$ ($\bar p\,p$)   & 181    & 352\\
$\sqrt{s}=630\,GeV$ ($\bar p\,p$)   & 19    & 78\\
$\sqrt{s}=1800\,GeV$ ($\bar p\,p$)   & 50    & 129\\
\hline
Total & 858 & 2738 \\
\hline
\end{tabular}
\end{center}
\caption{The quality of description of data on angular 
distributions.}
\label{tab2}
\end{table}

\begin{table}[h]
\begin{center}
\begin{tabular}{|l|l|l|}
\hline
Ref. & $\chi^2/d.o.f.$  &  kinematical range               \\
\hline
\cite{landshoff}   & Not presented    &
$23\,GeV\le \sqrt{s} \le 546\,GeV$\\
\cite{shelkovenko} & 2.0             &
$53\,GeV\le \sqrt{s} \le 630\,GeV\,,\;\;\;\;0 < -t \le 5\,GeV^2$\\
\cite{martynov}    & 2.4             &
$19\,GeV\le \sqrt{s} \le 1800\,GeV\,,\;\;\;\;0.1\,GeV^2 \le -t \le 14\,GeV^2$\\
\cite{povh}        & Not presented    &
$23\,GeV\le \sqrt{s} \le 546\,GeV$\\
\cite{bourrely}    & 4.3             &
$23\,GeV\le \sqrt{s} \le 1800\,GeV\,,\;\;\;\;0 < -t \le 6\,GeV^2$\\
\cite{islam}       & Not presented    &
$546\,GeV\le \sqrt{s} \le 1800\,GeV$\\
\cite{avila}       & 2.8             &
$23\,GeV\le \sqrt{s} \le 1800\,GeV\,,\;\;\;\;0.01\,GeV^2 \le -t \le 14\,GeV^2$\\
\cite{martynov2}   & 1.5             &
$6\,GeV\le \sqrt{s} \le 1800\,GeV\,,\;\;\;\;0.1\,GeV^2 \le -t \le 6\,GeV^2$\\
\cite{prokudin}    & 2.6             &
$23\,GeV\le \sqrt{s} \le 1800\,GeV\,,\;\;\;\;0.01\,GeV^2 \le -t \le 14\,GeV^2$\\
\hline
\end{tabular}
\end{center}
\caption{Some information on $\chi^2/d.o.f.$ over the data on 
elastic nucleon-nucleon scattering obtained by other authors.}
\label{tab3}
\end{table}

In the last two figures the ``soft'' pomeron and secondary trajectories 
and their slopes as 
functions of the transferred momentum squared are presented. 
Point out that although the intercepts of secondary reggeons 
in our scheme are 
higher than those ones from the rough linear approximation based on the 
appropriate resonance data the corresponding slopes 
turn out noticeably smaller than in the linear scheme. 
So, our approximations (\ref{fmeson}) to the true 
secondary trajectories in the scattering region may be laced in a 
smooth and monotonous manner 
with the corresponding Chew-Frautschi plots in the resonance 
region without any artificial behavior at low positive 
values of the argument (see fig.\ref{tra}). 

\section*{Conclusions}

Now we can discuss the correspondence of the Kwiecinski $q\bar q$-pole 
\cite{kwiecinski} to the secondary reggeons and the relevance 
of the Kirschner-Lipatov 
``hard'' pomeron pole \cite{kirschner} to elastic diffraction at accessible 
energies and small scattering angles. It is evident that the Kwiecinski 
asymptotic form 
(\ref{meson}) continued analytically in some way to the non-perturbative 
sector 
corresponds to the phenomenological secondary Regge trajectories giving a 
noticeable contribution to the observed elastic diffraction cross-sections. 
In other words, there exists a simple analytical way to connect their 
asymptotic perturbative behavior (\ref{meson}) with the Regge phenomenology 
in the non-perturbative sector. 

With the Kirschner-Lipatov ``hard'' pomeron the situation is 
quite different (see section ``The QCD pomeron''). Trying to apply the 
Kirschner-Lipatov pole to phenomenology we 
have an alternative: either to insist on the correspondence between the 
``soft'' pomeron trajectory and the Kirschner-Lipatov pole and so to accept that 
the trajectory is not monotonous in the Euclidean domain (in this case we come 
to the situation when we can not exploit its expression in form (\ref{hard}) for 
soft diffraction at any energies) or to assume that these poles are different 
ones (and so to presume that the Kirschner-Lipatov pole contribution to the 
eikonal is suppressed in the residue in the non-perturbative range of the 
argument value -- it must dominate in the diffraction sector at ultra-high 
energies (higher than $1.8\,TeV$) and also in 
the perturbative sector at accessible energies). The latter variant is more 
preferable from the analytical point of view but in both cases we are not able 
to use (\ref{hard}) in the Regge phenomenology of 
diffraction phenomena at accessible energies. As to (\ref{hardinter}) it 
gives even less reasons to use it, as explained above. 
The following picture seems to us reasonable. At ``very high'' 
$(-t)$ we deal with pure gluon exchanges, then, with $(-t)$ diminishing, we 
come to ``partially collectivized'' exchanges in the form of ``hard pomerons'' 
(\ref{hard}). The latter could be seen in gluon-gluon elastic scattering 
with colorless exchanges (``Mueller-Navelet jets'') with high $(-t)$. The 
hadron diffraction at $(-t)\le 3\,GeV^2$ is dominated by ``soft'' or 
non-perturbative pomerons (reggeons) which cannot be thought as composed 
of definite number of partons. Unfortunately, QCD-literature has no much to 
say in this case. 

\renewcommand{\labelenumi}{\theenumi)}

So, in the framework of the minimal Regge-eikonal model with using 
those parametrizations for Regge trajectories 
in which their asymptotic properties and fundamental 
nonlinearity are taken into account 
\begin{enumerate}
\item it was shown that the diffractive pattern for the 
elastic $\bar p\;p$ and $p\;p$ scattering at energies 
$23\,GeV < \sqrt{s} < 2\,TeV$ 
and scattering angles $-t < 3\,GeV^2$ is 
mainly formed by contribution of three Regge trajectories,
\item it was demonstrated (from both theoretical and 
phenomenological 
points of view) that we can not ignore the nonlinearity 
of the trajectories in the considered kinematical range,
\item the relevance of the Kwiecinski $q\bar q$-pole 
to the secondary reggeons and the impossibility to apply the 
Kirschner-Lipatov trajectory to elastic diffraction at 
accessible energies was argued.
\end{enumerate}

{\bf Acknowledgements:} The authors are indebted to V.V. Ezhela, 
A.V. Prokudin, R.A. Ryutin, S.M. Troshin and A.K. Likhoded for helpful 
discussions and useful criticism.

\newpage

\end{document}